\documentclass[aps,prl,reprint]{revtex4-1}
\usepackage{epsfig}
\usepackage{color}
\usepackage{amsmath}
\usepackage{amsfonts}
\usepackage{natbib}
\usepackage{notes2bib}

\newcommand{\degree}{^{\circ} }

\newcommand{\rb}{\mathbf{r}}

\newcommand{\avg}[1]{\left<#1\right>}
\newcommand{\len}[1]{\left|#1\right|}

\newcommand{\frho}{\hat{\rho}}

\newcommand{\kT}{\ensuremath{k_{\rm B}T}}

\begin{document}



\title{Lone Pair Rotational Dynamics in Solids}


\author{Richard C. Remsing}
\email[]{rick.remsing@rutgers.edu}
\affiliation{Department of Chemistry and Chemical Biology, Rutgers University, Piscataway, NJ 08854}
\author{Michael L. Klein}
\email[]{mike.klein@temple.edu}
\affiliation{Institute for Computational Molecular Science and Department of Chemistry, Temple University, Philadelphia, PA 19122}




\begin{abstract}
Traditional classifications of crystalline phases focus on nuclear degrees of freedom.
Through examination of both electronic and nuclear structure, we introduce the concept of an electronic plastic crystal.
Such a material is classified by crystalline nuclear structure, while localized electronic degrees of freedom --- here lone pairs ---
exhibit orientational motion at finite temperatures.
This orientational motion is an emergent phenomenon arising 
from the coupling between electronic structure and polarization fluctuations generated by collective motions, such as phonons.
Using \emph{ab initio} molecular dynamics simulations, we predict the existence of electronic plastic crystal motion
in halogen crystals and halide perovskites,
and suggest that such motion may be found in a broad range of solids with lone pair electrons.
Such fluctuations in the charge density should be observable, in principle via synchrotron scattering.
\end{abstract}


\maketitle

\raggedbottom

\begin{figure*}[t]
\begin{center}
\includegraphics[width=0.9\textwidth]{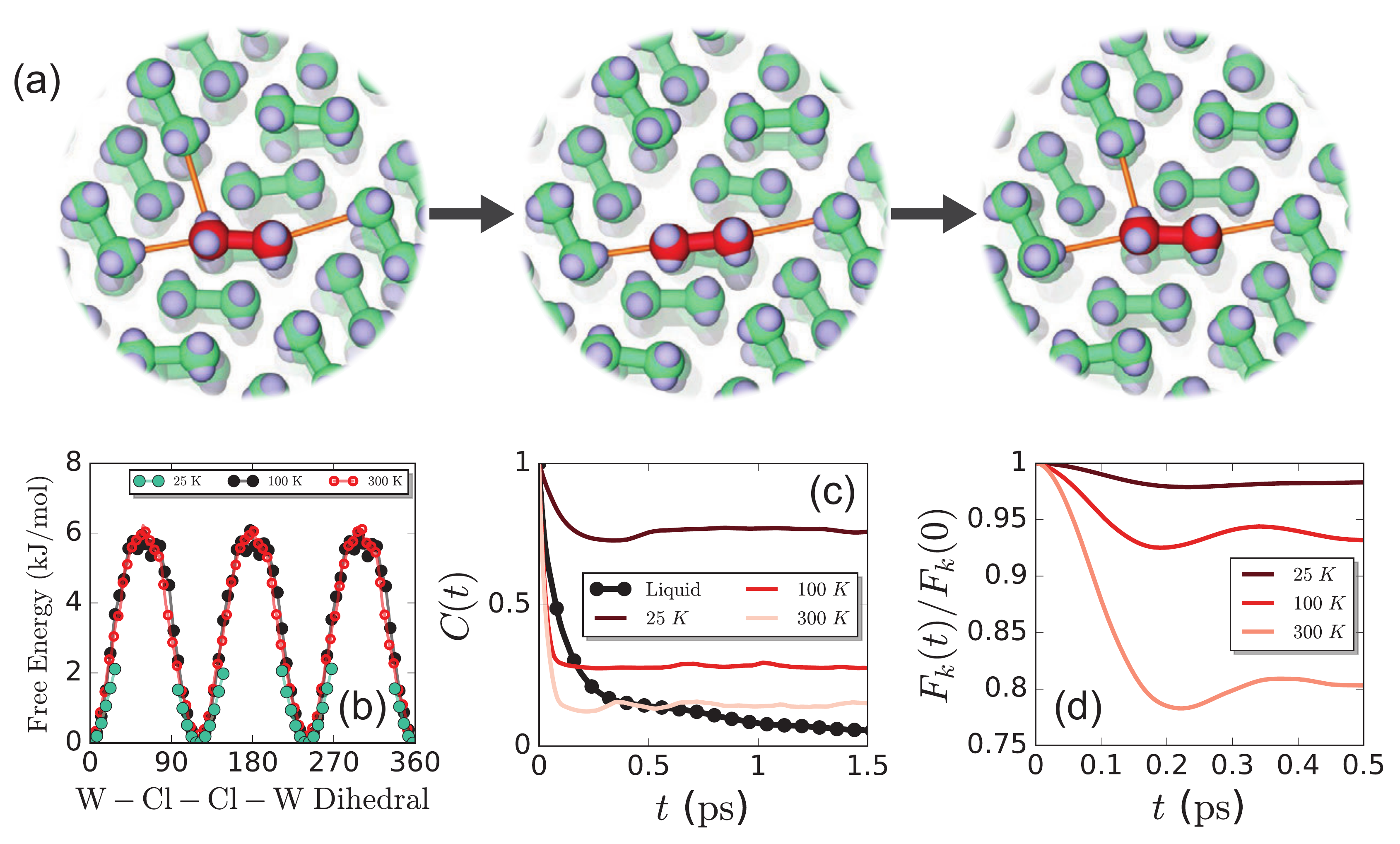}
\end{center}
\vspace{-10pt}
\caption
{
(a) Snapshots illustrating a lone pair rotation of 120$\degree$ in solid Cl$_2$ at $T=100$~K. 
Cl atoms are colored green and the maximally localized Wannier function centers (MLWFCs)
of the Cl lone electron pairs are shown as pale blue spheres. The Cl$_2$ molecule undergoing the rotation is colored red.
(b) Free energy as a function of the W-Cl-Cl-W dihedral angle, where W indicates the
center of a maximally localized Wannier function.
(c) Time correlation function for halogen bonds in liquid and solid Cl$_2$. Results for the solid
are shown for the same temperatures as in panel (b). 
(d) Intermediate scattering function, $F_k(t)$, determined according to Eq.~\ref{eq:fkt}, normalized by its value at $t=0$.
The value of $k=2.886$~\AA$^{-1}$ corresponds approximately to typical distances between lone pairs on opposite ends of the Cl$_2$ molecule.
}
\label{fig:cl2}
\end{figure*}

	Solids are phases of matter that break both translational and rotational symmetry, forming periodic atomic and/or molecular structures.
	In many molecular solids, increasing the temperature can lead to activation of rotational motion,
	such that the orientational structure of the activated modes becomes disordered while the translational symmetry
	is still broken and fixed on the periodic crystalline lattice~\cite{Staveley_AnnRevPC_1962,Klein_AnnRev_1985,Klein:ChemRev:1990,lynden1994translation}.
	These phases, characterized by long-ranged translational order and orientational disorder, are termed plastic crystals.
	Understanding the molecular details governing these orientationally disordered phases has led to profound insights into
	solid-state electrolytes~\cite{Klein:MolPhys:1995,MacFarlane:2016aa}, alkanes~\cite{Ungar:1983aa,PhysRevLett.58.698}, and fatty acid crystals~\cite{sirota1997remarks}, for example.
	In the classification of these phases, one focuses on the atomic (nuclear) structure of the materials.
	However, one might envision having similar correlations among electrons and nuclei,
	especially in systems with localized, lone pair electrons.
	In this work, we generalize the concept of a plastic crystal to electronic degrees of freedom
	and predict that solids can exhibit rotational lone pair dynamics as the temperature is increased
	while the nuclear degrees of freedom remain in the crystalline lattice structure.
	We detail this electronic plastic crystal motion in a model molecular crystal, Cl$_2$,
	and halide perovskites of the form ABX$_3$.
	This transition to an electronic plastic crystal phase may be significant to understanding
	reactivity, surface and phase behavior, photochemistry, and transport in materials.
To characterize the electronic plastic crystal motion, we perform ab initio molecular dynamics (AIMD) simulations
using CP2K and the \emph{QUICKSTEP} module~\cite{VandeVondele2005,VandeVondele2007}.
Simulations for Cl$_2$ systems followed our previous work~\cite{Remsing:JPCB:2019}.
For the perovskite simulations,
we employ the molecularly optimized (MOLOPT) Godecker-Teter-Hutter (GTH) double-$\zeta$ valence
single polarization short-ranged (DZVP-MOLOPT-SR-GTH) basis set~\cite{VandeVondele2007}
and the GTH-PADE pseudopotential~\cite{Goedecker1996}
to represent the core electrons.
The valence electrons were treated explicitly, using the PBE~\cite{PBE} functional as implemented in CP2K
with a plane wave cutoff of 400~Ry, in order to connect to earlier work on similar systems~\cite{Fabini:2016aa}.
We first equilibrated each system to the desired temperature using a Nos\'{e}-Hoover
thermostat chain of length three~\cite{Nose_1984,Nose_1984b} with a timestep of 1.0~fs.
Systems were then equilibrated in the microcanonical (NVE) ensemble,
before gathering statistics in the NVE ensemble over at least 4~ps of production simulation time.
The coordinates of the maximally localized Wannier function centers (MLWFCs) were obtained using CP2K, minimizing the MLWF spreads
according to the formulation of Ref.~\onlinecite{PhysRevB.61.10040}.
	We first focus on solid Cl$_2$ as a model molecular solid that exhibits rotational lone electron pair dynamics.
	Diatomic chlorine forms a single covalent Cl-Cl bond and the remaining six electrons of each Cl form
	three $sp^3$ hybridized lone pairs, as illustrated by the MLWFCs~\cite{RevModPhys.84.1419} in Figure~\ref{fig:cl2}a.
	In addition, the Cl$_2$ molecule has electron deficient, $\sigma$-hole regions located along the bond axis at the end of each
	Cl, as well as between each lone pair~\cite{Desiraju:AccChemRes,Cavallo_2016,Remsing:JPCB:2019}.
	The unique orthorhombic crystal structures of the halogens Cl$_2$, Br$_2$, and I$_2$ are stabilized by halogen bonds,
	directional electrostatic attractions between (negative) lone electron pairs and these (positive) $\sigma$-holes~\cite{Cavallo_2016,I2-2014,Remsing:JPCB:2019}.
	%

	%
	Despite the presence of halogen bonds, we find that, at high enough temperatures, the lone electron pairs of Cl$_2$
	rotate about the Cl-Cl bond axis.
	This rotational motion can be observed in Figure~\ref{fig:cl2}a,
	where we show three snapshots of the lone pair MLWFCs along a simulation trajectory.
	The combined electronic/nuclear structure of Cl$_2$ is reminiscent of the nuclear structure of ethane (C$_2$H$_6$),
	which exhibits rotational motion of hydrogen atoms in its plastic crystal phase that resemble that of the Cl$_2$ lone pairs shown here~\cite{Straty_JCP_1976,eggers1975plastic}.
	To further the analogy to ethane,
	we can characterize the Cl$_2$ electronic plastic crystal motion by defining a lone pair-Cl-Cl-lone pair (W-Cl-Cl-W)
	dihedral angle, $\phi$, and examining its statistics.
	The effective free energy landscape governing lone pair rotations, $\Delta F(\phi)$, is then given by
	$\Delta F(\phi)=-\kT \ln P(\phi)$, where $\kT$ is the product of Boltzmann's constant and the temperature
	and $P(\phi)$ is the probability distribution of the dihedral angle $\phi$ observed in the simulation.
	This free energy is shown in Figure~\ref{fig:cl2}b for three temperatures, one below the electronic plastic crystal transition (25~K),
	one above the transition and below melting (100~K), and a superheated state (300~K).
	The three-fold symmetry of $\Delta F(\phi)$ arises from the symmetry of the Cl$_2$ lone pairs.
	At low temperatures, thermal fluctuations (in the form of activated phonon modes) are not large enough to activate orientational
	motion of the lone pairs, and the free energy can only be computed near the minima; barriers are not traversed under unbiased sampling.
	At 100~K and 300~K, electronic plastic crystal motion is observed in the solid, lone pairs readily rotate between ground states,
	and the free energy barrier with a temperature-independent height of $\Delta F^\ddag\approx6$~kJ/mol is sampled.

	The rotational motion of lone pairs breaks halogen bonds in order to cross the free energy barrier,
	and then reforms halogen bonds upon completing a rotation and returning to a free energy minimum.
	The lone pair rotational motion will therefore show up in time-dependent quantifications of halogen bond dynamics.
	We characterize the dynamics of halogen bonds through the time correlation function (TCF) $C(t)=\avg{h(t)h(0)}/\avg{h}$,
	where $h(t)=1$ if a halogen bond between two Cl atoms is intact at time $t$, and $h(t)=0$ otherwise~\cite{Remsing:JPCB:2019,LuzarChandler,Luzar_2000}.
	The halogen bond TCF is shown in Figure~\ref{fig:cl2}c for the same three states discussed above.
	We observe a significant change in the form of $C(t)$ as the rotational motion of lone pairs is activated;
	$C(t)$ plateaus at a much lower value and the initial decay is much faster at high $T$.
	This faster decay of $C(t)$ is consistent with the increased rotational motion of lone pairs in solid Cl$_2$,
	which transiently break halogen bonds between neighboring molecules.
	The disruption of halogen bonding increases with temperature and may play an important role in
	melting for example, wherein the transient weakening of intermolecular interactions by lone pair rotations
	could make it easier to nucleate a liquid phase than would be the case if lone pairs orientations were fixed.
	We note that the description of halogen bonding can depend sensitively on the choice of density functional,
	with charge transfer playing a significant role in some cases~\cite{Klein_ChargeTransfer,MHG_XB}.
	However, we expect our findings to be qualitatively insensitive to these subtleties, with changes in the halogen bond strength
	leading to shifts in the onset temperature for lone pair dynamics being the dominant effect.

\begin{figure}[t]
\begin{center}
\includegraphics[width=0.49\textwidth]{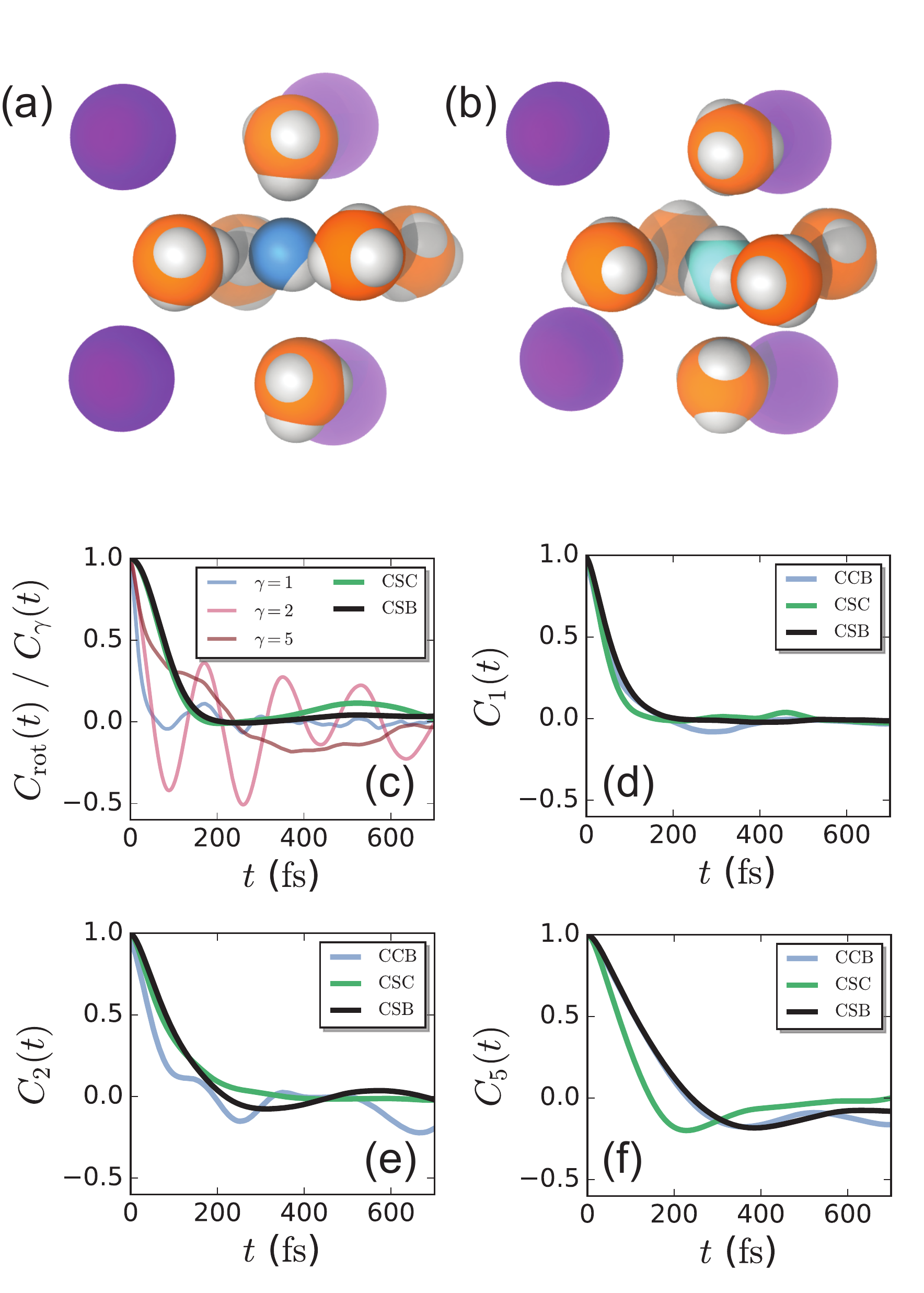}
\end{center}
\vspace{-10pt}
\caption
{
(a,b) Snapshots highlighting the coordination environment around a central (a) Sn in CsSnBr$_3$ and (b)
Ca in CsCaBr$_3$. Cs are colored purple, Br are orange, Sn is blue, and Ca is cyan. In both panels,
maximally localized Wannier function centers (MLWFCs) of Br, Sn, and Ca are shown as gray spheres.
(c) Rotational time correlation functions for B-sites as defined in Eq.~\ref{eq:rotcor} and Eq.~\ref{eq:cgamma} for the Sn-lone pair MLWFC dipole moment in CSC/CSB
and the Ca MLWFCs in CCB, respectively. The Ca correlation functions are indicated by their value of $\gamma$.
(d,e,f) Rotational correlation functions for X-sites (Cl or Br) defined in Eq.~\ref{eq:cgamma} for $\gamma=1,2,5$.
}
\label{fig:perov}
\end{figure}

	Experimentally, electron dynamics can be probed through inelastic scattering~\cite{Chergui:2017aa,Ultrafast_SwissFEL}. 
	Within this context, the key observable is the intermediate scattering function
	\begin{equation}
	F_k(t)=\avg{\frho_k(t)\frho_{-k}(0)},
	\end{equation}
	where $\frho_k(t)$ is the Fourier transform of the electron density, $\rho(r,t)$, at time $t$.
	Exact computation of $F_k(t)$ would require solving the time-dependent Schr\"{o}dinger equation
	to monitor the quantum dynamics of the electrons in the system.
	However, we can approximate $F_k(t)$ using the results from our AIMD simulations, where the dynamics
	are contained only in the nuclear motion, and the electron density is constrained to lie at the ground state
	in each nuclear configuration.
	Within this level of approximation, it is not necessary to work within the basis of eigenstates of the Hamiltonian,
	and we can determine the electron density in each configuration (at each time) from the MLWFs~\cite{RevModPhys.84.1419}. 
	If the shape of the MLWFs is rigid, as is true to a good approximation for Cl$_2$, we can further approximate
	the electron density as a convolution of a \emph{time-independent} shape function, $f(r)$, and the density of
	MLWFCs, $\rho^{\rm C}(r,t)$, such that the intermediate scattering function is
	\begin{equation}
	F_k(t)\approx \hat{f}_k \avg{\frho^{\rm C}_k(t) \frho^{\rm C}_k(0)}.
	\label{eq:fkt}
	\end{equation}
	Thus, for rigid MLWFs and Born-Oppenheimer AIMD, the electron dynamics are contained in the trajectories
	of the MLWFCs, drastically simplifying the estimation of $F_k(t)$.
	The intermediate scattering function, $F_k(t)$, is shown for select values of $k$ and a range of temperatures in Figure~\ref{fig:cl2}d.
	In agreement with the behavior of $C(t)$, the scattering function decays more rapidly when lone pair rotational motion is present,
	and exhibits almost no decay or features beyond the initial transient below the electronic plastic crystal transition.
	This result suggests that inelastic scattering-based probes of electron dynamics may be able to uncover the existence
	electronic plastic phases in materials.
	Additionally, one might also envision probing electronic dynamics indirectly through NMR relaxation and chemical shift anisotropy measurements~\cite{Obermyer_1973,bryce2006solid,Cl-CSA}.
	Dynamical motion of lone pairs is not limited to molecular solids.
	We also find significant
	lone pair rotational motion in the ABX$_3$ halide perovskites CsSnCl$_3$ (CSC), CsSnBr$_3$ (CSB), and CsCaBr$_3$ (CCB), at 400~K.
	In these systems, the nuclei-MLWFC structure of Cl/Br and Ca are topologically analogous to
	methane molecules, in the same way that the Cl$_2$ molecule's electronic structure was akin to ethane.
	Therefore, we can expect that the rotational dynamics of Cl/Br and Ca in these perovskites may resemble
	those of the plastic phases of methane and other systems containing tetrahedral molecules~\cite{NIJMAN1977188,SPRIK1980411,James_Keenan,Klein_Tetrahedral,lynden1994translation}.
	The Sn atom has a single lone pair, forming a nuclei-lone pair dipole when the lone pair MLWFC
	is off-center, i.e. when the MLWFC-Sn bond length is greater than zero. 
	Such off-centering occurs in the cubic phase studied here, as evidenced in previous work~\cite{C7SC01429E,Fabini:2016aa}
	and by the snapshot in Fig.~\ref{fig:perov}a.
	In contrast, Ca has a symmetric lone pair structure in this perovskite, as indicated
	by the snapshots illustrating typical MLWFC structures in Fig.~\ref{fig:perov}b.

	%
	We characterize the lone pair motion in the halide perovskites through rotational TCFs.
	For the Sn-lone pair dipole motion, we compute
	\begin{equation}
	C_{\rm rot}(t) = \avg{P_2(\mathbf{\mu}(t)\cdot\mathbf{\mu}(0))},
	\label{eq:rotcor}
	\end{equation}
	where $\mu(t)$ is the Sn-MLWFC dipole vector at time $t$ and $P_2(x)$ is the 
	second order Legendre polynomial.
	For Cl/Br and Ca atoms, we compute the TCF of tetrahedral rotor functions, $M_\gamma$, of order $l=3$,
	following previous work on ionic crystals with tetrahedral ions~\cite{Klein_Tetrahedral}.
	Here, $\gamma$ labels the $(2l+1)$ functions for each $l$.
	Due to the cubic symmetry of the crystal, we need to only consider three representative functions,
	$M_1 = 3\sqrt{3}/4 \sum_{i=1}^4 x_i y_i z_i$, $M_2=3\sqrt{5}/40 \sum_{i=1}^4 (5x_i^3 - 3 x_i r_i^2)$,
	and $M_5=3\sqrt{3}/8 \sum_{i=1}^4 x_i (y_i^2-z_i^2)$,
	where $\rb_i=(x_i,y_i,z_i)$ is a unit vector along nuclei-MLWFC bond $i$, and $r_i=\len{\rb_i}$.
	We then examine the motion of Cl/Br and Ca MLWFCs through the TCFs
	\begin{equation}
	C_{\gamma}(t) = \avg{\delta M_\gamma(t) \delta M_\gamma(0)}/\avg{\delta M_\gamma^2(0)},
	\label{eq:cgamma}
	\end{equation}
	where $\delta M_\gamma(t) = M_\gamma(t)-\avg{M_\gamma}$.
	The TCFs $C_{\rm rot}(t)$ and $C_\gamma(t)$ are shown in Fig.~\ref{fig:perov}c-f
	and suggest that lone pair rotational motion occurs on rapid, sub-picosecond timescales. 
	The rotational timescale of the B-site ion (ABX$_3$) is roughly the same as that of the X-site ions,
	highlighting the interplay of lone pair rotational dynamics that gives rise to dynamic off-centering of Sn
	observed in CsSnBr$_3$ and similar materials~\cite{PhysRevB.67.125111,Fabini:2016aa}.
	Moreover, the timescale for MLWFC rotational motion is in agreement with that identified for
	local polar fluctuations in similar perovskites~\cite{PhysRevLett.118.136001}. 
	These polar fluctuations were linked to Br face expansion and Cs head-to-head motions.
	Our results suggest that the coupled rotational dynamics of the Sn-lone pair dipole and the halide electron density
	drive these motions, and ultimately local polarity fluctuations and dynamic disorder in cubic lead halide perovskites.

	We find only subtle differences between the rotational timescales of the B-site and X-site among the three perovskites studied.
	In particular, we find that rotations in CSC may be slightly faster than in CSB, most likely due to the larger polarizability of Br leading
	to stronger X-B ion-ion interactions. 
	 The observed fast decay of orientational correlations suggests the ability of these perovskites to rapidly respond to the addition of a charge to the lattice,
	 either through a charged defect or photoexcited charge carriers. 
	 In this context, one is concerned with the solid-state solvation dynamics of the system~\cite{Madigan_2003,Delor_2017,Cotts_2017}.
	 Recent work has highlighted the utility of applying concepts from liquid-state solvation theory to polaron formation in halide perovskites~\cite{Zhu:2015aa,Zhu1409,Miyatae1701469,Guo:2019aa}.
	 Within this liquid-state context, solvation dynamics are characterized by the time dependence of the interaction energy between the charge and its environment following introduction of the solute charge~\cite{Stratt:JPC:1996,Thompson_2011}.
	 Maroncelli and coworkers have shown that in many cases, including solvation in a dipolar lattice,
	 that such a response function can be approximated reasonably well by a power-law scaling of the dipole rotational TCF,
	 where the power is proportional to the dipole density~\cite{Maroncelli:1993aa,Maroncelli:BrownianLattice}. 
	 Thus, within the accuracy of this model, a rapid decay of $C_{\rm rot}(t)$ implies fast solvation dynamics within halide perovskites.
	 These findings are in agreement with the high efficiency of CSB to separate photoexcited charged carriers.
	 If the dipoles are not highly correlated, nanoscale polarization (or solid-state solvation) in response
	 to a charge carrier is not significantly affected by polarization to other charge carriers.
	 Thus, the solvation environment around one charge carrier does not `see' that around another,
	 and the interactions between charge carriers are efficiently screened, especially on timescales longer
	 than the short solvation dynamics timescale implied by $C_{\rm rot}(t)$.
	 This efficient solvation can also be expected from the large dielectric constant of CSB ($\sim67$), while that for CCB
	 is much lower ($\sim17$).
	 We note, however, that the difference in the dielectric constants of CSB and CCB is not expected to originate from the timescale for dipole/polarization fluctuations,
	 because the rotational times of Br in each crystal, as well as those of Sn and Ca, are approximately the same, Fig.~\ref{fig:perov}c-f.
	 Instead, our results suggest that polarization fluctuations occur on the same timescale in CSB and CCB, however,
	 the dipole moment that is fluctuating in CSB is of a much larger magnitude, which gives rise to the large dielectric constant
	 and ultimately the higher efficiency of CSB as a photovoltaic material.

\begin{acknowledgements}
This work was supported as part of the
Center for Complex Materials from First Principles (CCM), an Energy Frontier
Research Center funded by the U.S. Department of Energy, Office of Science, Basic Energy
Sciences under Award \#DE-SC0012575.
Computational resources were supported in part by the National Science Foundation
through major research instrumentation grant number 1625061
and by the US ARL under contract number W911NF-16-2-0189.
\end{acknowledgements}

\bibliography{LPD}

\end{document}